# Lateral electrodeposition of MoS$_2$ semiconductors over an insulator

*Nema M. Abdelazim, Yasir J. Noori[*], Shibin Thomas, Victoria K. Greenacre, Yisong Han, Danielle E. Smith, Giacomo Piana, Nikolay Zhelev, Andrew L. Hector, Richard Beanland, Gillian Reid, Philip N. Bartlett, Cornelis H. de Groot[*]*

Dr. N. M. Abdelazim, Dr. Y. J. Noori, Dr. G. Piana, Prof. C. H. de Groot
Electronics and Computer Science, University of Southampton, SO17 1BJ, Southampton, UK
E-mail: y.j.noori@southampton.ac.uk; chdg@ecs.soton.ac.uk

Dr. S. Thomas, Dr. V. K. Greenacre, Dr. D. E. Smith, N. Zhelev, Prof. A. L. Hector, Prof. G. Reid, Prof P. N. Bartlett
School of Chemistry, University of Southampton, SO17 1BJ, Southampton, UK

Dr Y. Han, Prof R. Beanland
Department of Physics, University of Warwick, CV4 7AL, Coventry, UK

N. M. Abdelazim, Y. J. Noori and S. Thomas have contributed equally to this work.

Keywords: (Electrodeposition, 2D-materials, transition metal dichalcogenides, MoS$_2$, photodetectors)

**Abstract**

Developing novel techniques for depositing transition metal dichalcogenides is crucial for the industrial adoption of 2D materials in optoelectronics. In this work, the lateral growth of molybdenum disulfide (MoS$_2$) over an insulating surface is demonstrated using electrochemical deposition. By fabricating a new type of microelectrodes, MoS$_2$ 2D films grown from TiN electrodes across opposite sides have been connected over an insulating substrate, hence, forming a lateral device structure through only one lithography and deposition step. Using a variety of characterization techniques, the growth rate of MoS$_2$ has been shown to be highly anisotropic with lateral to vertical growth ratios exceeding 20-fold. Electronic and photo-response measurements on the device structures demonstrate that the electrodeposited MoS$_2$ layers behave like semiconductors, confirming their potential for photodetection applications. This lateral growth technique paves the way towards room temperature, scalable and site-selective production of various transition metal dichalcogenides and their lateral heterostructures for 2D materials-based fabricated devices.

# 1. Introduction

Molybdenum disulfide is a two-dimensional (2D) transition metal dichalcogenide (TMDC) semiconductor material that has been used as a building block in demonstrations of high on/off ratio transistors, ultrasensitive photodetectors and sensors.[1–4] Some of these demonstrations were implemented for wearable applications by exploiting the material's exceptional robustness and flexibility.[5,6] However, there remain major obstacles that hinder the industrial adoption of $MoS_2$ and other 2D TMDC materials. The most challenging obstacle has been finding an industrially compatible method that enables the production of these materials on a mass scale. We have recently demonstrated that electrodeposition is a potentially viable method for solving this challenge.[7,8] Electrodeposition offers important advantages in 2D material production over other methods such as chemical vapor deposition (CVD),[9,10] sputtering[11] or atomic layer deposition (ALD).[12] Electrodeposition is not a line-of-sight deposition method as material growth occurs at electrical contacts and is controlled by electrical potential or current.[13] It can hence be utilized to deposit materials over three dimensional surfaces including patterned nanostructures of high aspect ratios.[14–17] In addition, electrodeposition is usually performed at room temperature, avoiding harsh environments such as plasma or extremely high temperatures, which can damage pre-existing materials on the substrate, such as graphene electrodes.[7]

However, there is an important limitation with electrodeposition that need to be overcome. This method requires an electrically conductive surface from which materials are traditionally grown vertically.[18] Depositing a semiconductor material on a conductor provides a low resistance current path in planar (opto-) electronic devices such as transistors and photodetectors, thus limiting the use of electrodeposition traditionally to certain vertical device structures or metal interconnects (through the dual damascene process).[19] Prior to the work described herein, this limitation has been a drawback, specifically for developing 2D material based devices where planar structures that exploit the unique 2D properties of the material are the "natural" route forward.[20–23]

Creating innovative techniques to electrodeposit planar 2D materials over non-conducting surfaces would solve this limitation and open new routes where the insulator base can be utilized, such as in transistor gating. In the early 1990s, Nishizawa et al. described the electrochemical growth of poly(pyrrole) from an electrode out over the surface of an insulator, controlled by modification of the insulator surface.[24–26] More recently, Kobayashi et al. have showed that the lateral growth rate of Au from an Au electrode film can be increased via the surface treatment of $SiO_2$.[27] However, these are relatively rare examples. Furthermore, these demonstrations required surface treatment, and were limited to certain materials which severally restrict the scope of device applications that this technique can be employed for.

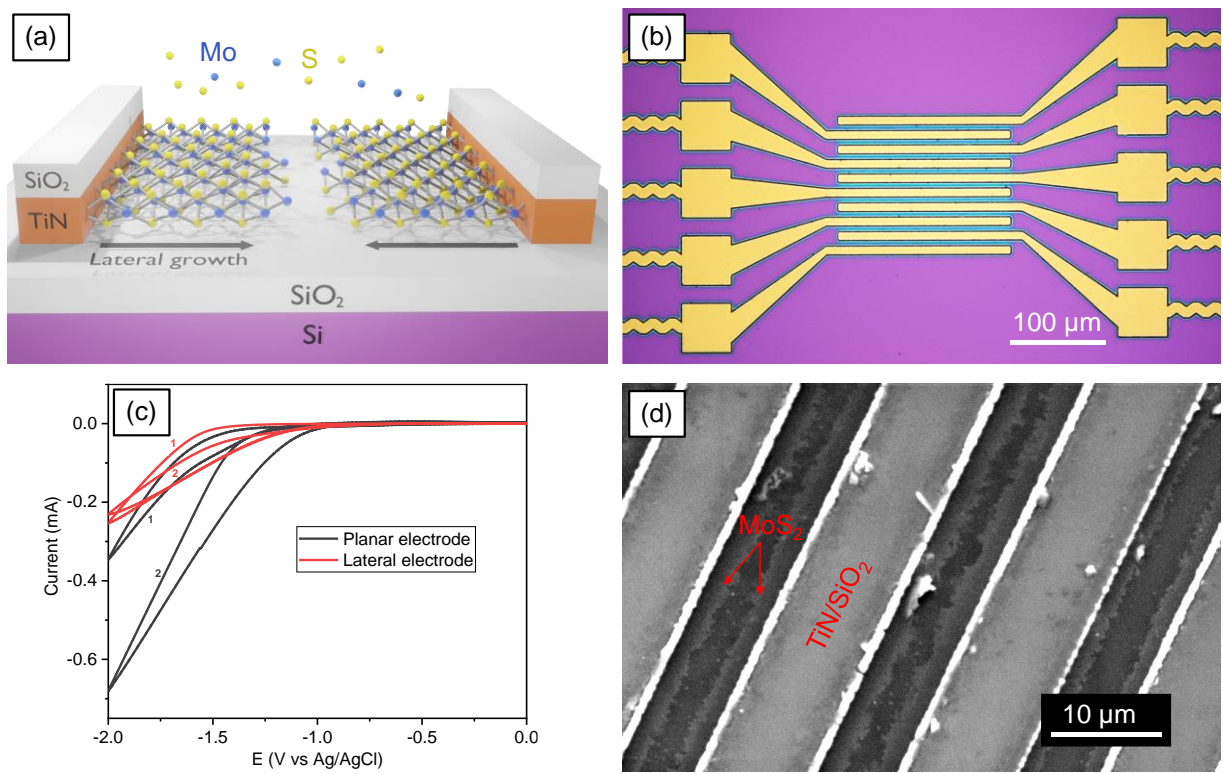

**Figure 1.** (a) An illustration of the concept of this work showing TiN electrodes that are top covered with a SiO$_2$ insulator with a TMDC MoS$_2$ film growing laterally on the SiO$_2$/Si substrate. (b) A microscope image of the fabricated electrode structure showing 10 adjacent electrodes each connected to a pad for electrical contact. The light blue colour surrounding the edges of the electrodes is laterally grown MoS$_2$ (c) Cyclic voltammetry electrodeposition scans comparison between a large-area planar TiN electrode (diameter 4 mm) and one of the TiN lateral growth electrodes. (d) An SEM image showing four TiN electrodes and three micro-gaps that are partially filled with laterally grown MoS$_2$ films via electrodeposition.

In this paper, we exploited the lateral growth technique for producing 2D TMDC semiconductors and their devices. We demonstrated strong anisotropy in the lateral growth rate of ultra-thin binary MoS$_2$ compounds via the electrodeposition over insulating substrates. By fabricating a new microelectrode structure that has its top coated with an insulator but its side edges exposed, we were able to restrict the growth of MoS$_2$ to favor the lateral growth direction for several micrometers, starting from these thin edges. The anisotropic growth along the lateral direction is demonstrated by transmission electron microscopy to be particularly attractive for two dimensional materials due to their unique planar layered structures. Using various material characterization techniques, we have shown that the lateral growth rate of MoS$_2$ across the insulator is 20-fold larger than its vertical growth rate. Furthermore, we have discovered that the lateral growth rate is linear, which allowed us to controllably connect two MoS$_2$ films grown from electrodes positioned on opposite sides across an insulator. This approach enabled us to realize a MoS$_2$ based device structure via a single lithography step. Finally, we demonstrated that our laterally grown MoS$_2$ films are semiconductors by developing an array of photodetector devices.

## 2. Fabrication and Electrodeposition

### 2.1. Lateral growth electrodes fabrication

An illustration of the concept of our work on the lateral electrodeposition of MoS$_2$ is shown in Figure 1a. The TiN electrodes were fabricated on a SiO$_2$/Si substrate using a single photolithographic step as shown in Figure S1. First a negative photoresist was spin-coated on a wafer and then UV exposed using a photolithographic mask to form the desired pattern. TiN and SiO$_2$ were then consecutively sputtered on the wafer and a lifted-off was performed to form the lateral growth electrodes. The thickness of the sputtered TiN layer is approximately 100 nm. The double layer lift-off was executed to minimize the number of fabrication steps, and to eliminate the possibility of misalignment between the two sputtered layers that could result from performing multiple lithographic steps. A microscope image of the fabricated electrodes is shown in Figure 1b. The wafer was then diced into smaller chips prior to MoS$_2$ electrodeposition. An illustration of the chip layout is shown in supplementary Figure S2. After MoS$_2$ deposition, each chip is cleaved into two halves through the cleave zone to disconnect the global electrode and to create individual devices.

While all of the electrodes fabricated in this work were based on TiN, this fabrication technique should in principle be applicable to electrodes based on other materials, such as Au. However, challenges may arise due to alloying between the electrode material with the deposit, especially during annealing at high temperatures. The advantage of using TiN is that it does not alloy easily due to its Ti-N covalent bonds in the lattice.

### 2.2. Electrodeposition of MoS$_2$

The electrodeposition experiments were performed in CH$_2$Cl$_2$ solvent using [N$^n$Bu$_4$]$_2$[MoS$_4$], which was synthesized in-house to function as a single source precursor, providing both the Mo and S.[7,8] In contrast to [NH$_4$]$_2$[MoS$_4$], which has been used in aqueous solvents, [N$^n$Bu$_4$]$_2$[MoS$_4$] has excellent solubility in CH$_2$Cl$_2$. The electrodeposition solution also included [N$^n$Bu$_4$]Cl as the supporting electrolyte and trimethylammonium chloride [NHMe$_3$]Cl as the proton source, which is necessary to remove the excess S during deposition of MoS$_2$ from [MoS$_4$]$^{2-}$.[8] The electrodeposition experiments were all performed within a glovebox equipped with a dry nitrogen circulation system to minimize moisture and ensure that oxygen levels are maintained below 10 ppm. The depositions were performed using a three electrode electrochemical cell with a Pt gauze as counter electrode and a Ag/AgCl (0.1 M [N$^n$Bu$_4$]Cl in CH$_2$Cl$_2$) reference electrode.

Cyclic voltammetry (CV) scans were performed to study the electrochemical behavior of the electrolyte with the lateral growth electrodes. Figure 1c shows the comparison of CVs recorded on a TiN fabricated electrode and a 4 mm diameter planar TiN electrode. The CVs are obtained by sweeping the voltage from 0 to -2.0 V in the cathodic scan and then -2.0 to 0 V in the anodic scan. On the first cycle there is evidence of a nucleation process in both cases with the current being increased on the return and on subsequent scans. This occurs because the deposited MoS$_2$ functions as a catalyst for the reduction of the NHMe$_3^+$ which results in hydrogen evolution. We recently reported a detailed investigation on the electrochemical processes during the CV of the same electrolyte system employing an electrochemical quartz crystal microbalance (EQCM) measurements.[8] The CVs displayed in Figure 1c clearly show evidence of the cathodic reduction of [MoS$_4$]$^{2-}$ ions to MoS$_2$. It is clear that the CV recorded from the lateral TiN electrodes is comparable, in terms of electrochemical processes at the interface, to the CV measured on the planar TiN electrode, indicating that the electro-reduction mechanism of the [MoS$_4$]$^{2-}$ ions remains the same irrespective of the type of TiN electrode used.

The lateral growth of MoS$_2$ was achieved through potentiostatic electrodeposition by applying -1.0 V and varying the deposition time to achieve the desired lateral growth length. The electrochemical reactions at the working and counter electrodes respectively, are as follows

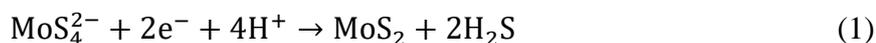

$$\text{MoS}_4^{2-} + 2e^- + 4H^+ \rightarrow \text{MoS}_2 + 2H_2S \tag{1}$$

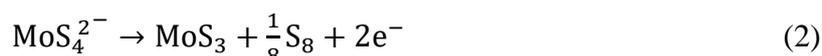

$$\text{MoS}_4^{2-} \rightarrow \text{MoS}_3 + \frac{1}{8}S_8 + 2e^- \tag{2}$$

Following electrodeposition, the sample was rinsed with CH$_2$Cl$_2$ and left to dry inside the glove box. The as-deposited MoS$_2$ film is amorphous, as X-ray diffraction on thicker films indicate, although with some short-range ordering present as evidenced by the preferential lateral growth.[8,28–30] An annealing step was performed to crystallize the film. The sample was annealed using a tube furnace that was set initially to 100 ºC for 10 min and then 500 ºC for 2 h at 0.1 mbar. Annealing the films was performed within a sulfur rich environment made by placing 0.1 g of sulfur powder together with the sample inside the quartz tube within the tube furnace. To prevent distillation of volatile sulfur from the film and cause it to become understoichiometric, a sulfur rich environment was used. Thermal annealing was found to substantially improve the crystallinity of the film as evidenced by Raman Spectroscopy.[7,8] It also removed excess sulfur from the film.

## 3. Results and Discussions

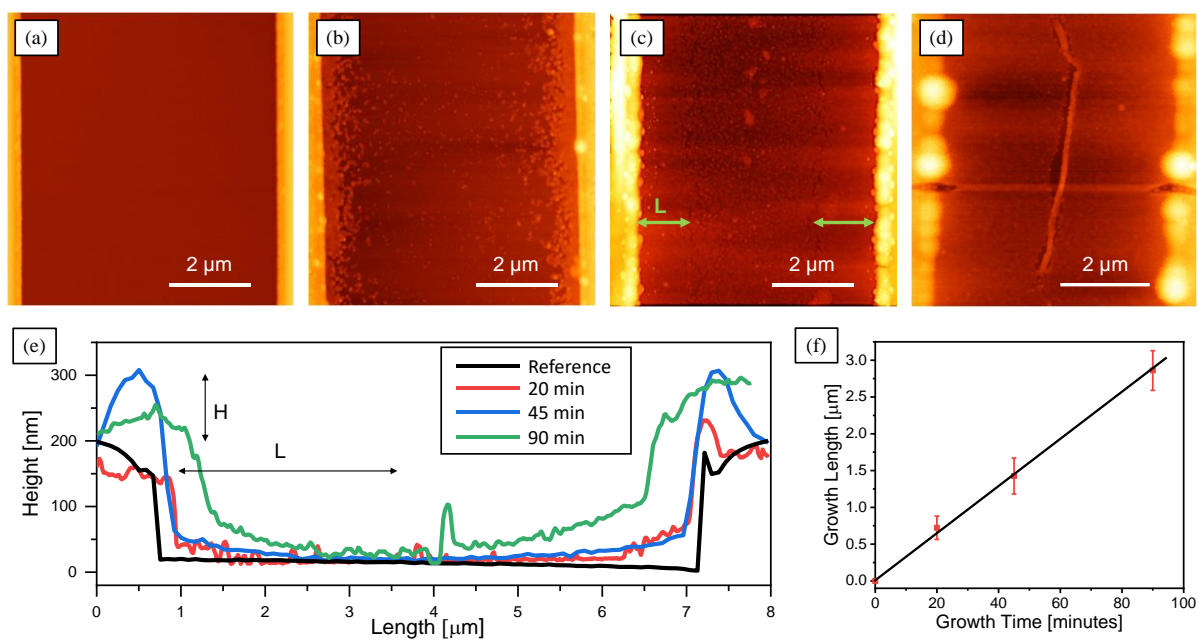

**Figure 2.** AFM images of (a) reference sample and (b-d) laterally grown films deposited for (b) 20 min, (c) 45 min and (d) 90 min after anneal. e) Topography line profiles collected from images (a-d). The figure shows indications of the vertical growth (H) and lateral growth (L) that were considered for analysis in this work. f) A linear fit of the growth length vs time obtained from the three samples.

### 3.1. Microscopy

On applying an electrodeposition potential to the electrode array, the material starts to grow from the side edges of each electrode. Throughout this works, the deposition time was chosen to be either 20, 45 or 90 minutes. Figure 1b shows an optical microscope image of patterned electrodes with laterally grown $MoS_2$ films depicted by the light blue borders surrounding the yellow electrodes. Figure 1d shows a scanning electron microscope (SEM) image of a laterally grown $MoS_2$ film from the edges of TiN electrodes. The contrast difference between the laterally grown $MoS_2$ films and the $SiO_2$ covered substrate underneath can be clearly observed. Supplementary Figure S3 (a-c) shows SEM images of laterally grown films deposited for 20, 45 and 90 minutes. By observing the electrodeposition length with time, we were able to allow the laterally grown films from adjacent electrodes to contact in the middle region by electrodepositing for 90 minutes.

Figure 2 (a-d) shows atomic force microscopy (AFM) images performed across the growth regions between two adjacent electrodes from a pristine substrate and ones with laterally grown $MoS_2$ films. Typical line profiles for each film are shown in Figure 2e. After 20 minutes of deposition (Figure 2b), the AFM images show clear initial growth of the film laterally from both sides of the electrodes. Upon increasing the deposition time to 45 min (Figure 2c), the film near to the electrodes were found to be smoother and continuous. However, it is clear from the figure that 45 min is not sufficient to allow the films grown from opposite sides to contact each other. Figure 2d shows an AFM image of a film grown for 90 minutes. In the latter deposition, the two films were found to meet in the center between the electrodes above the $SiO_2$ substrate, causing one film to grow over the other, resulting in a small peak in the AFM profile in the middle region.

**3.2. Films Thickness and Length**

Figure 2f shows that the lateral growth length scales linearly with the deposition time. An average growth rate of 33 ± 6 nm/min was extracted. The error bars represent the maximum and minimum observed growth lengths. The lateral growth length measurements were recorded on the films that were grown out of the upper most electrode from the ten-electrode array shown in Figure 1b. This was chosen to prevent connecting layers from introducing uncertainties in the length measurements. Each experiment with a particular deposition time was replicated to ensure experimental reproducibility. The average lengths and uncertainties were calculated over ten AFM measurements from every sample. To quantify the thickness and length of the grown films, topography profiles were acquired from five lines across AFM images corresponding to films grown for 20, 45 and 90 min as shown in supplementary Figure S4a. The average values of the vertical (height) and lateral growths (length) were extracted for each film. The average heights above the edge of the electrodes ($H_{av}$), lengths ($L_{av}$) and their corresponding maximum ($H_{max}$ and $L_{max}$) values were recorded in Table 1 in the supplementary information. Figure S4b shows the $L_{av}$ and $H_{av}$ measurements taken from five different lines across the 20, 45 and 90 minutes grown films. By comparing the values of $H$ versus $W$, we found that the lateral growth rate is approximately 20 times larger than the vertical growth. In these experiments, the electrodes thickness was fixed to 100 nm. We think that using thinner electrodes will increase the growth anisotropy, due to surface effects, compared to thicker ones but will not suppress the vertical growth rate completely.

As the electrodeposition time is increased, the vertical growth of the material becomes more prominent, especially near the electrodes where the film growth starts. We observed that the vertically grown materials can be thick enough to rise above the fabricated $TiN/SiO_2$ electrodes and start to grow laterally over the top insulated electrodes. However, it is clear that the vertical growth rate of $MoS_2$ is much slower than the lateral growth rate.

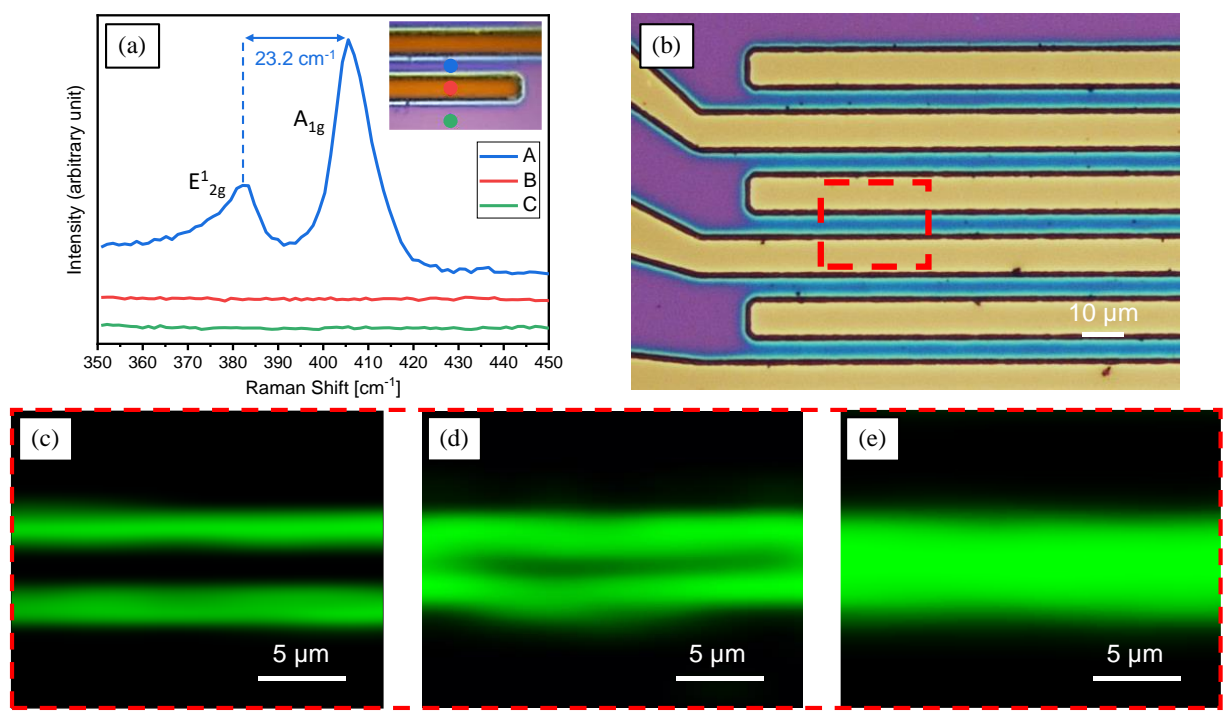

**Figure 3.** a) A stack of Raman spectra showing that the MoS$_2$ signal is solely observed from the laterally grown films with no Raman signatures from areas away from the electrodes or at the centre above them. b) An optical microscope image of a series of fabricated TiN micro-electrodes with laterally grown MoS$_2$ films (90 minutes) that are showing as light blue borders surrounding the electrodes. c-e) Raman maps of areas corresponding to that of the red box in (b) between adjacent electrodes for films grown for 20, 45 and 90 minutes, respectively. The maps show a clear increase in the lateral length of the MoS$_2$ films with increasing the deposition time.

### 3.3 Raman Spectroscopy

Raman spectroscopy is a commonly used technique in characterizing different physical properties of TMDC 2D materials. We used this method to confirm the presence of the MoS$_2$ films and qualitatively study their degree of crystallinity and lateral growth length. All measurements were performed using a 532 nm excitation laser at room temperature. A 50x objective lens was used to focus the excitation laser to a ~1 µm spot diameter and simultaneously collect the emitted light. Figure 3a shows three Raman spectra collected from different locations around the electrodes using a sample with a MoS$_2$ film grown for 45 minutes. Point A is in the center between the two electrodes, point B is above one of the electrodes, point C is on a blank SiO$_2$/Si region of the substrate, away from the electrodes. These locations are shown in the inset of the figure. The MoS$_2$ Raman signature corresponding to the E$^1_{2g}$ and the A$_{1g}$ peaks was only found at point A, demonstrating that there is no deposition at the center above the top insulated TiN electrodes nor on the SiO$_2$ substrate away from the electrodes. The central positions of the E$^1_{2g}$ and A$_{1g}$ peaks are 382.5 cm$^{-1}$ and 405.6 cm$^{-1}$, respectively. Energy dispersive X-ray (EDX) spectroscopy spectra are shown in Figure S5. Due to the large overlap between the Mo and S electron emission lines (~2.3 keV), EDX is not a reliable technique to measure the elemental composition. It was therefore only used to identify the nature of the materials at different regions of the sample, utilizing the much higher image resolution offered by the SEM compared to optical techniques. The EDX spectra show Mo and S signatures solely above the laterally grown films, confirming the conclusion of the Raman images. In previous works, we characterized the material composition of $MoS_2$ films electrodeposited via similar electrochemical environments using wavelength dispersive x-ray spectroscopy (WDX) and x-ray photoelectron spectroscopy (XPS).[7,8] These results have shown that the Mo:S ratio is ~1:2. Raman mappings were performed on selected areas as exemplified in Figure 3b of 20, 45 and 90 minutes grown films using the $A_{1g}$ peak signal intensity and a step size of ~1 μm, see Figure 3 (c-e). The maps show the evolution of the lateral growth as the electrodeposition time is increased, causing the films grown from opposite sides to slowly merge after approximately 90 min deposition. However, the fact that the laser spot is ~1 μm means that the exact distribution of the material cannot be precisely defined through this method. The combined results of AFM, SEM, and Raman mapping clearly confirm the lateral growth of $MoS_2$ across the $SiO_2$ insulator after starting from the thin edges of the TiN electrodes.

### 3.4 Transmission Electron Microscopy

Figure 4a shows a high-resolution cross section SEM image of a laterally grown film. The cross-section was taken after performing a lamella process using a focused ion beam (FIB). The image shows the $MoS_2$ thin film grown over the $SiO_2$/Si substrate and covered by Pt and C protection layers that were deposited during the lamella process. It can be seen that the film gets thinner further away from the electrode towards the left of the image. Annular dark field (ADF) TEM imaging was then used to reveal the nanocrystalline structure of $MoS_2$ layers. The ADF mode allows distinguishing $MoS_2$ from its surrounding $SiO_2$ and Pt. Figure 4b shows a laterally grown film which was deposited for 90 minutes. We noticed that the stacked layers are generally aligned in the lateral growth direction perpendicular to the plane of the substrate. In a film grown for 45 minutes, presented in Figure 4 (c and d), TEM imaging reveals that the film is much thinner and where the layers are more ordered. This suggest that the amount of ordering correlates inversely with the thickness of the films. The layer-to-layer distance of the ordered stack shown in Figure 4c was measured to be $0.7 \pm 0.1$ nm, which is in agreement with literature values of $MoS_2$ layers spacing measurements.[31] Conventional X-ray diffraction is not suitable to test the crystallinity of these laterally grown films due to their small area and thickness. However, X-ray diffraction performed on large area and thick deposition attempts performed in similar environments showed the appearance of a crystallisation peak corresponding to (002) plane of 2H-MoS$_2$ following film annealing.[8] The TEM results of the laterally grown films presented here show that the films are polycrystalline with layer sizes in the order of tens of nanometers.

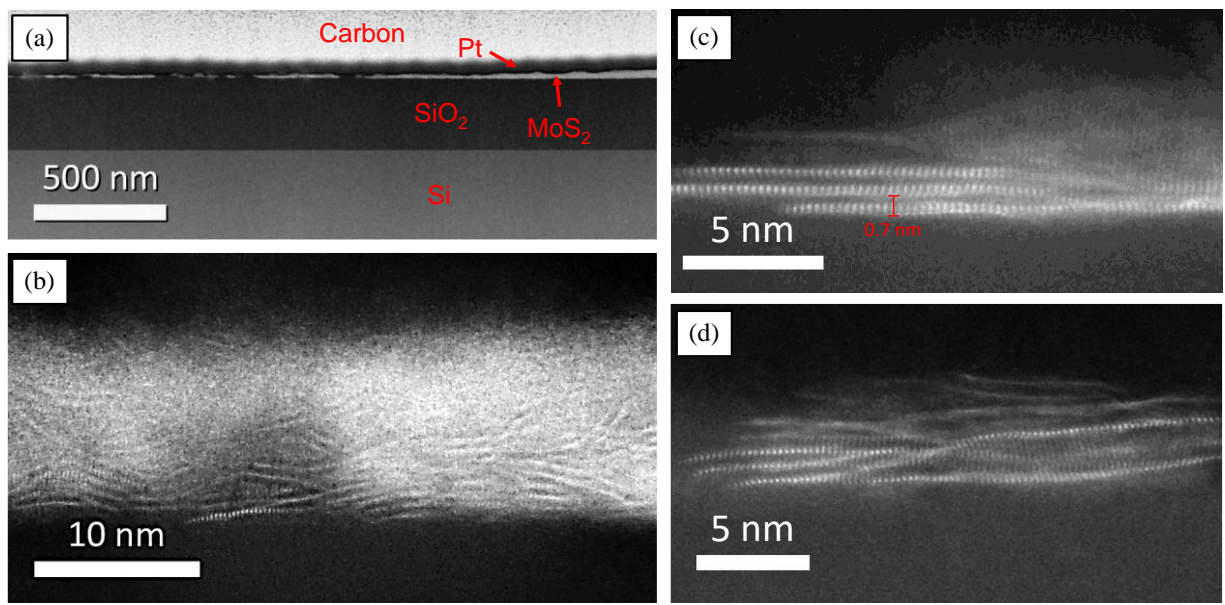

**Figure 4.** a) A high-resolution SEM image of a cross section of a region between two electrodes with a laterally grown MoS$_2$ film. The film was coated with Pt and C protective layers (b) A magnification of the MoS$_2$ film shown in (a) taken via TEM that clearly shows the layered nature of the film growing preferentially in the horizontal direction along the surface of the wafer. The film was taken from a sample that was grown for 90 minutes. c-d) TEM images of a few layers of MoS$_2$ taken at a thin region from a film that was grown for 45 minutes.

### 3.5 Photoresponsivity of laterally grown MoS$_2$ devices

Electrical characterization of the laterally grown films was performed by connecting adjacent lateral growth electrodes to a semiconductor device analyzer. Figure 5a shows current-voltage sweeps taken from films that were grown for 20, 45, and 90 minutes. The displayed curves were taken from different films that were grown in multiple repeat experiments. The electrical connection between the electrodes for the 20 min sample is open, while that for the 45 minutes sample, has resistance exceeding 500 MΩ. The electrical resistance was measured to be significantly lower for the 90 minutes sample after the laterally grown films from both sides have connected in the middle. The inset of the figure shows ohmic behavior with a resistance of ~1 MΩ. Using the length and width of the channel and the estimated average thickness of the films (~ 60 nm), we extract the room temperature resistivity of the 90 minutes grown films as 115 Ω.cm. This resistivity is in the range of earlier reported MoS$_2$ films.[32]

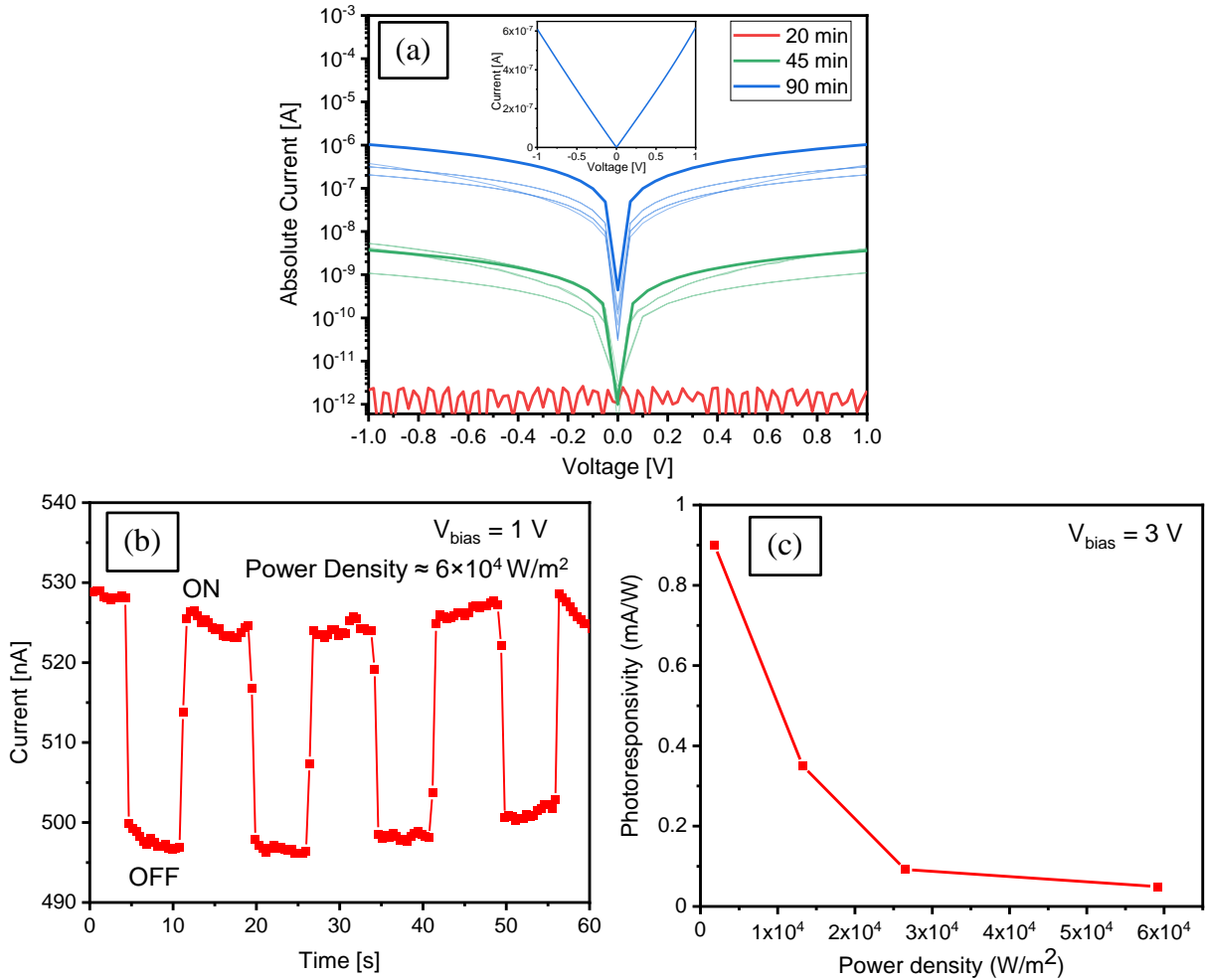

**Figure 5.** a) Multiple Current-Voltage sweeps from MoS$_2$ devices grown multiple times for 20, 45 and 90 minutes showing an open circuit after 20 min film growth and a great increase in current from 45 min growth to 90 minutes. b) photo-illumination cycles of a sample showing the switching induced photocurrent with a switching laser source. c) The change in photoresponsivity as the power density of the excitation laser is changed.

The photoresponsivity of the MoS$_2$ films was tested at room temperature in air, using a 532 nm laser source. The aim of this experiment was to test the film's semiconductor optical absorption and induced current. The tests were performed by applying a bias voltage on two adjacent electrodes and the current was measured through the device in the dark and under laser illumination. The laser light was coupled to the chip through a fiber and a long distance 20x objective lens, reducing the spot size to a circular spot with a diameter of roughly 100 µm. Figure 5b presents the photocurrent induced in the material due to a pulsing laser. The applied bias, $V_{bias}$ = 1 V and the laser power density was calculated to be around 6×10$^4$ Wm$^{-2}$. The photoresponsivity of the film was calculated using equation (3) below:

$$Photoresponsivity = \frac{I_{ph}}{P_{laser}} \qquad (3)$$

where $I_{ph}$ is the photoinduced current. $I_{ph}$ can be calculated by subtracting the device current under illumination from the dark current $I_{ph} = I_{light} - I_{dark}$ and $P_{laser}$ is the incident laser power. Several measurements were made on different chips containing devices of films grown for 90 minutes. The maximum photoresponsivity recorded was 0.9 mA W$^{-1}$. We then performed photoresponsivity measurements at varying laser power density as shown in Figure 5c.

It was found that the photoresponsivity reduces at higher laser power density, indicating carrier saturation in the material. Whilst this photoresponsivity is higher than our previously values reported from the electrodeposition method,[7] it is still lower than previously reported from $MoS_2$ films made via CVD and mechanical exfoliation[33,34]. This is expected to be due to the lower material crystallinity produced from electrodeposition in comparison to the other aforementioned methods. Our future work will include using a similar structure to develop phototransistor devices using the bottom $SiO_2$ as a back-gate dielectric.[2,9]

## 4. Conclusions

We have demonstrated the lateral electrodeposition of transition metal dichalcogenide (TMDC) films over an insulating substrate for optical and electrical measurements. The $MoS_2$ is deposited directly onto patterned electrodes using a non-aqueous electrodeposition method. The nature of the deposition is such that the material growth occurs from the side edges of 100 nm thick TiN electrodes. The electrodeposition duration determines the extent of the material's lateral growth. By fabricating adjacent electrodes and exploiting the preference of the material to grow laterally, we have shown the ability to grow films from opposite sides towards each other until they connect in the middle. The deposited films were characterized using a variety of techniques including AFM, Raman, SEM, EDX, and TEM. The lateral growth length was found to increase linearly with the deposition time with an average rate of ~ 33 ± 6 nm/min. This is around 20 folds faster than the material's vertical growth. Electrical and photoresponsivity measurements of the $MoS_2$ films showed that the material is semiconducting. Our work provides an innovative, efficient and scalable method for the lateral growth of 2D materials and promotes their applications in next-generation electronic and optoelectronic devices. This paves the way towards future possibilities such as electrodepositing different TMDCs to form lateral heterostructures of 2D materials, creating novel p-n-p junction in a single electrodeposition experiment.[35,36]

**Acknowledgements**

The research work reported in this article was financially supported by the Engineering and Physical Sciences Research Council (EPSRC) through the research grant EP/P025137/1 (2D layered transition metal dichalcogenide semiconductors via non-aqueous electrodeposition) and the programme grant EP/N035437/1 (ADEPT - Advanced Devices by ElectroPlaTing).**Conflicts of Interest**

The authors declare no conflicts of interest.

**References**

[1]    X.-F. Wang, H. Tian, Y. Liu, S. Shen, Z. Yan, N. Deng, Y. Yang, T.-L. Ren, *ACS Nano* **2019**, *13*, 2205.

[2]    Z. Yin, H. Li, H. Li, L. Jiang, Y. Shi, Y. Sun, G. Lu, Q. Zhang, X. Chen, H. Zhang, *ACS Nano* **2012**, *6*, 74.

[3]    H. S. Nalwa, *RSC Adv.* **2020**, *10*, 30529.

[4]    R. Kumar, W. Zheng, X. Liu, J. Zhang, M. Kumar, *Adv. Mater. Technol.* **2020**, *5*, 1901062.

[5]    S. Bertolazzi, J. Brivio, A. Kis, *ACS Nano* **2011**, *5*, 9703.

[6]    N. Li, Q. Wang, C. Shen, Z. Wei, H. Yu, J. Zhao, X. Lu, G. Wang, C. He, L. Xie, J. Zhu, L. Du, R. Yang, D. Shi, G. Zhang, *Nat. Electron.* **2020**, *3*, 711.

[7]    Y. J. Noori, S. Thomas, S. Ramadan, D. E. Smith, V. K. Greenacre, N. Abdelazim, Y. Han, R. Beanland, A. L. Hector, N. Klein, G. Reid, P. N. Bartlett, C. H. Kees de Groot, *ACS Appl. Mater. Interfaces* **2020**, *12*, 49786.

[8]    S. Thomas, D. E. Smith, V. K. Greenacre, Y. J. Noori, A. L. Hector, C. H. (Kees) de Groot, G. Reid, P. N. Bartlett, *J. Electrochem. Soc.* **2020**, *167*, 106511.

[9]    W. Zhang, J.-K. Huang, C.-H. Chen, Y.-H. Chang, Y.-J. Cheng, L.-J. Li, *Adv. Mater.* **2013**, *25*, 3456.

[10]   J. Chen, W. Tang, B. Tian, B. Liu, X. Zhao, Y. Liu, T. Ren, W. Liu, D. Geng, H. Y. Jeong, H. S. Shin, W. Zhou, K. P. Loh, *Adv. Sci.* **2016**, *3*, 1500033.

[11]   J. Tao, J. Chai, X. Lu, L. M. Wong, T. I. Wong, J. Pan, Q. Xiong, D. Chi, S. Wang, *Nanoscale* **2015**, *7*, 2497.

[12]   X. Tang, N. Reckinger, O. Poncelet, P. Louette, F. Ureña, H. Idrissi, S. Turner, D. Cabosart, J.-F. Colomer, J.-P. Raskin, B. Hackens, L. A. Francis, *Sci. Rep.* **2015**, *5*, 13523.

[13]   A. A. Ojo, I. M. Dharmadasa, *Electroplating of Semiconductor Materials for Applications in Large Area Electronics: A Review*, Vol. 8, **2018**.

[14]   N. J. Gerein, J. A. Haber, *J. Phys. Chem. B* **2005**, *109*, 17372.

[15]   H.-F. Wang, C. Tang, Q. Zhang, *Nano Today* **2019**, *25*, 27.

[16] R. Huang, G. P. Kissling, R. Kashtiban, Y. J. Noori, K. Cicvarić, W. Zhang, A. L. Hector, R. Beanland, D. C. Smith, G. Reid, P. N. Bartlett, C. H. (Kees) de Groot, *Faraday Discuss.* **2019**, *213*, 339.

[17] P. N. Bartlett, S. L. Benjamin, C. H. (Kees) de Groot, A. L. Hector, R. Huang, A. Jolleys, G. P. Kissling, W. Levason, S. J. Pearce, G. Reid, Y. Wang, *Mater. Horizons* **2015**, *2*, 420.

[18] J. L. Hudson, T. T. Tsotsis, *Chem. Eng. Sci.* **1994**, *49*, 1493.

[19] P. C. Andricacos, C. Uzoh, J. O. Dukovic, J. Horkans, H. Deligianni, *IBM J. Res. Dev.* **1998**, *42*, 567.

[20] A. Sharma, R. Mahlouji, L. Wu, M. A. Verheijen, V. Vandalon, S. Balasubramanyam, J. P. Hofmann, W. M. M. (Erwin) Kessels, A. A. Bol, *Nanotechnology* **2020**, *31*, 255603.

[21] M. Chhowalla, D. Jena, H. Zhang, *Nat. Rev. Mater.* **2016**, *1*, 16052.

[22] F. Giannazzo, G. Greco, F. Roccaforte, S. S. Sonde, *Crystals* **2018**, *8*, 70.

[23] Y. Liu, N. O. Weiss, X. Duan, H.-C. Cheng, Y. Huang, X. Duan, *Nat. Rev. Mater.* **2016**, *1*, 16042.

[24] M. Nishizawa, M. Shibuya, T. Sawaguchi, T. Matsue, I. Uchida, *J. Phys. Chem.* **1991**, *95*, 9042.

[25] M. Nishizawa, Y. Miwa, T. Matsue, I. Uchida, *J. Electrochem. Soc.* **1993**, *140*, 1650.

[26] M. Nishizawa, T. Kamiya, H. Nozaki, H. Kaji, *Langmuir* **2007**, *23*, 8304.

[27] C. Kobayashi, M. Saito, T. Homma, *Electrochim. Acta* **2012**, *74*, 235.

[28] A. Albu-Yaron, *Electrochem. Solid-State Lett.* **1999**, *2*, 627.

[29] R. Chaabani, A. Lamouchi, B. Mari, R. Chtourou, *Mater. Res. Express* **2019**, *6*, 115902.

[30] A. Lamouchi, I. Ben Assaker, R. Chtourou, *J. Mater. Sci.* **2017**, *52*, 4635.

[31] J. Xiao, M. Long, X. Li, Q. Zhang, H. Xu, K. S. Chan, *J. Phys. Condens. Matter* **2014**, *26*, 405302.

[32] G. Eda, H. Yamaguchi, D. Voiry, T. Fujita, M. Chen, M. Chhowalla, *Nano Lett.* **2011**, *11*, 5111.

[33] G. Wu, X. Wang, Y. Chen, Z. Wang, H. Shen, T. Lin, W. Hu, J. Wang, S. Zhang, X. Meng, J. Chu, *Nanotechnology* **2018**, *29*, 485204.

[34] N. Perea-López, Z. Lin, N. R. Pradhan, A. Iñiguez-Rábago, A. Laura Elías, A. McCreary, J. Lou, P. M. Ajayan, H. Terrones, L. Balicas, M. Terrones, *2D Mater.* **2014**, *1*, 11004.

[35]  X. Duan, C. Wang, J. C. Shaw, R. Cheng, Y. Chen, H. Li, X. Wu, Y. Tang, Q. Zhang, A. Pan, J. Jiang, R. Yu, Y. Huang, X. Duan, *Nat. Nanotechnol.* **2014**, *9*, 1024.

[36]  J. Wang, Z. Li, H. Chen, G. Deng, X. Niu, *Nano-Micro Lett.* **2019**, *11*, 48.